\documentclass[prl,aps,epsf,amsfonts,floats,twocolumn,amssymb,amsmath,groupedaddress,showpacs,floatfix,nofootinbib]{revtex4}
\usepackage{graphicx}
\usepackage{epsf}

\usepackage[]{latexsym}
\usepackage{bm}

\newcommand{\be}{\begin{equation}}\newcommand{\ee}{\end{equation}}
\newcommand{\bea}{\begin{eqnarray}}\newcommand{\eea}{\end{eqnarray}}
\newcommand{\brr}{\begin{array}}\newcommand{\err}{\end{array}}
\newcommand{\bit}{\begin{itemize}}\newcommand{\eit}{\end{itemize}}
\newcommand{\ben}{\begin{enumerate}}\newcommand{\een}{\end{enumerate}}

\newcommand{\ba}{\begin{array}}
\newcommand{\ea}{\end{array}}

\def\lf{\left}

\def\non{\nonumber}

\def\ri{\right}

\def\1{{_{1}}}\def\2{{_{2}}}

\def\noHe0{:\;\!\!\;\!\!:H_e(0):\;\!\!\;\!\!:}
\def\noHm0{:\;\!\!\;\!\!:H_\mu(0):\;\!\!\;\!\!:}

\def\lf{\left}

\def\non{\nonumber}

\def\ri{\right}

\def\1{{_{1}}}\def\2{{_{2}}}

\begin{document}

\title{Vacuum condensates as a mechanism of spontaneous supersymmetry breaking}

\author{ Antonio Capolupo\footnote{Corresponding author:   e-mail address: capolupo@sa.infn.it (A.Capolupo).}}

\author{ Marco Di Mauro}

\affiliation{ Dipartimento di Fisica E.R.Caianiello, Universit\'a di Salerno, and INFN Gruppo collegato di Salerno, Fisciano (SA) - 84084, Italy}


\begin{abstract}

A possible mechanism for the spontaneous breaking of SUSY, based on the presence of vacuum condensates, is reviewed. Such a mechanism could occur in many physical examples, both at the fundamental and emergent level, and would be formally analogous to spontaneous SUSY breaking at finite temperature in the TFD formalism, in which case it can be applied as well. A possible experimental setup for detecting such a breaking through  measurement of the Anandan-Aharonov invariants associated with vacuum condensates in an optical lattice model is proposed.

\end{abstract}

\maketitle

\section{Introduction}

Supersymmetry (SUSY) \cite{Golfand:1971iw,Wess:1992cp} has had a huge impact   on contemporary physics, not only from the purely theoretical and mathematical point of view, but also from the phenomenological and experimental ones. This despite the absence, up to now, of any clear experimental signature of its existence at the fundamental level. The main reason for this is that to date SUSY provides the best available explanation the gauge hierarchy problem of the Standard Model \cite{Weinberg:1976zw}, as well as providing candidates for dark matter and improving the situation of the dark energy issue (which however is still far from settled). In a few years the situation may radically change due to the results coming from the LHC, but it is a fact that if SUSY exists at a fundamental level, it must be broken, either spontaneously or explicitly, since otherwise the superpartners of the known particles would be degenerate with the latter and thus would have been observed long ago. For this reason over the years there has been a lot of activity concerning SUSY breaking, especially, but non only, focusing on the spontaneous breaking case (see e.g. \cite{Witten:1981nf,Shadmi:1999jy} and references therein).

Besides fundamental SUSY, an interesting possibility, both on its own right and as a way to experimentally test ideas on SUSY and its breaking in the near future, is \emph{emergent} SUSY. Namely, it is possible that a condensed matter system may display SUSY at low energies, which may or may not be spontaneously broken. In particular, relativistic supersymmetric theories could be simulated with cold atom systems in optical lattices \cite{Yue Yang}. In what follows we shall describe a mechanism for SUSY breaking, based on vacuum condensates, which may be valid both at a fundamental and at an emergent level \cite{Capolupo:2012vf,Capolupo:2013af}. The latter possibility also suggests ways to investigate this mechanism in table top experiments. 

The idea is to exploit the formal analogy between thermal field theory in its Thermo--Field Dynamics (TFD) formulation \cite{Takahashi:1974zn} and different physical phenomena characterized by vacuum condensates similar to those appearing in the thermal context \cite{Schwinger:1951nm}--\cite{Blasone:2002jv}. As in the thermal case SUSY is spontaneously broken (see below), we expect that this happens in the same way also in these other phenomena. A possible experiment involving the measurement of the Anandan--Aharonov invariant associated with the vacuum condensate is also described.

Before explicitly stating our conjecture, let us briefly recall how SUSY is spontaneously broken in TFD.

It is well known that SUSY is spontaneously broken at any finite temperature \cite{Buchholz:1997mf,Das:1997gg}, the fundamental reason being the different statistical behavior of bosons and fermions. Finite temperature physics can be formulated in a way which is equivalent to the standard ensemble based picture, but which emphasizes the appearance of vacuum condensates. This formalism goes under the name of Thermo-Field Dynamics \cite{Takahashi:1974zn}. In this formalism vacuum condensates in the thermal ground state are conveniently described by means of Bogoliubov transformations, and thermal effects are encoded in the appearance of a new vacuum which is unitarily inequivalent to the zero temperature one. Thermal averages are then just vacuum expectation values with respect to this new vacuum. In the standard picture \cite{Buchholz:1997mf,Das:1997gg}, SUSY breaking is due to the fact that it is not possible to write down thermal averages in a way consistent with SUSY, while in the TFD picture it is due to the fact that the new vacuum acquires a nonvanishing energy density. This picture thus links thermal breaking of SUSY to the standard description of SUSY breaking, whose order parameter is precisely the vacuum energy density \cite{Witten:1981nf,Shadmi:1999jy}. This last fact is as well known a straightforward consequence of the SUSY algebra: if the vacuum is not invariant under SUSY transformations, i.e. $Q_{\alpha}|0\rangle \neq 0$, then (here $Q$ is the supercharge that generates SUSY transformations, $H$ is the Hamiltonian of the theory and $C$ is the charge conjugation matrix)
\bea
\langle 0|H|0\rangle = \frac{1}{8} \,\langle 0| \textrm{Tr}(C\gamma^0\lf[Q,Q\ri]_{+})|0\rangle \neq  0,
\eea
while of course if the vacuum is invariant then $\langle 0|H|0\rangle=0$. Physically, this is due to the fact that the zero point energies of fermions and bosons cancel; schematically
\bea\label{Canc}
H=H_{\psi}+H_B &\sim&\sum_{\mathbf{k},i} \,\left\{  \omega_{k,i}^\psi\lf( N_{k,i}^\psi - \frac{1}{2}  \ri)+ \right.\\  \nonumber &&+\left.\omega_{k,i}^B \lf( N_{k,i}^B + \frac{1}{2}  \ri) \right\},
\eea
and in a supersymmetric theory $\omega_{k,i}^\psi=\omega_{k,i}^B \equiv\omega_{k,i}$. In the case of TFD, the condensates which are present in the thermal vacuum lift the vacuum energy. Such a lift is not canceled in a supersymmetric theory, thereby triggering the spontaneous breaking of SUSY.

The point is that the formalism of Bogoliubov transformations \cite{Bogolyubov:1947zz}, on which this vacuum condensate based description of thermal physics is founded, is quite universal, and describes vacuum condensates in a host of different quantum field theoretical (QFT) phenomena at various length scales, from fundamental to emergent models \cite{Umezawa:1993yq}. Examples of such phenomena include quantum fields in external fields, such as Schwinger \cite{Schwinger:1951nm} and Unruh \cite{Unruh:1976db} effects, examples from condensed matter physics such as the BCS theory of superconductivity \cite{Bardeen:1957mv} and graphene \cite{Iorio:2010pv}, mixing in particle physics \cite{Blasone:2001du,Blasone:2002jv}\footnote{In the case of mixing the situation is slightly different, in that the Bogoliubov transformation is nested in a unitary transformation of the fields, however this does not qualitatively change what we will say.}.
This leads to the conjecture that in all these cases, when a supersymmetric extension is possible at the classical level, vacuum condensates lift the vacuum energy, thereby spontaneously breaking SUSY \cite{Capolupo:2012vf,Capolupo:2013af}.
We give some evidence for this conjecture by considering the free Wess--Zumino model \cite{Wess:1973kz}. Despite its simplicity, this simple picture should give a good qualitative understanding of the vacuum of more complicated systems.

Considering the range of the phenomena described by this picture, this mechanism may occur at a fundamental level, for example triggered by particle mixing, as proposed in \cite{Capolupo:2010ek, Mavromatos:2010ni}, or at an emergent level. The first possibility is very interesting from a phenomenological point of view and may be used for model building, while the latter possibility suggests, as said, the possibility of conceiving experimental measurements of the vacuum energy due to the condensates, therefore corroborating our conjecture \cite{Capolupo:2012vf}. This will be also the object of the present paper, in which the possibility of probing thermal spontaneous SUSY breaking through geometric invariants \cite{Capolupo:2013ica} will be explored. To be specific, the relevant quantity is the Anandan-Aharonov invariant \cite{Anandan:1990fq}, which has been shown to be a feature of phenomena characterized by vacuum condensates \cite{Capolupo:2013fasi}.

A few comments are in order. First of all, since vacuum condensates are a genuine field theoretical and nonperturbative phenomenon, this kind of SUSY breaking can occur only in QFT, and it is nonperturbative in nature (consistently with the fact that if SUSY is unbroken at tree level, it can only be broken at the nonperturbative level \cite{Witten:1981nf}). Second, while in what follows we shall give evidence for our conjecture in a simple case, we do not address the issue of the dynamical origin of that breaking or, which is the same, of the origin of the vacuum condensates, which depends on the specific details of the phenomena under study, and which in some cases such as mixing is to date unknown. The effective description of condensates in terms of Bogoliubov transformations is instead universal (besides being technically straightforward), since the form of this transformations is always the same, the details of the specific case being encoded in the coefficients. This means that our discussion will be necessarily qualitative, while a more quantitative approach will need dealing with the complexities of the dynamics on the various cases. In particular, the computation of quantities such as the scale of the breaking and mass differences between superpartners lies beyond the scope if the present paper. Also, we do not address the issue of the Goldstone fermion associated with the breaking.
This issue, as well as the detailed study of some specific case, are left for some future publication.

\section{Vacuum condensate and SUSY breaking}

As a model of the supersymmetric extension of any of the above systems, we consider a situation in which SUSY is preserved at the lagrangian level and study the vacuum condensation effects. These are described by a Bogoliubov transformation acting simultaneously, and with the same parameters, on the bosonic and on the fermionic degrees of freedom. This is required in order not to break SUSY explicitly. We conjecture that in such a situation, SUSY is spontaneously broken by the appearance of vacuum condensates. In the present section,  we collect some basic facts about Bogoliubov transformations in QFT (see e.g. \cite{Umezawa:1993yq}), then we prove in a simple case that vacuum condensates do shift the vacuum energy.

The modes of any boson  (fermion) field are described by a set of ladder operators $a_{\mathbf{k}}$, whose canonical (anti) commutation relations (CCRs) are:
$
[a_{\mathbf{k}}, a^{\dagger}_{\mathbf{p}}]_{\pm}=\delta^{3}(\mathbf{k}-\mathbf{p})\,,
$
with $-$ for bosons  and $+$ for fermions and all other (anti) commutators vanishing. The vacuum $|0\rangle$ is defined by $a_{\mathbf{k}}|0\rangle$, and a Fock space is built out of it by acting with the creation operators $a^{\dagger}_{\mathbf{k}}$.

A generic Bogoliubov transformation has the form:
\bea\label{Atilde}
\tilde{a}_{\mathbf{k}}(\xi) = U_{\mathbf{k}} \, a_{\mathbf{k}} - V_{\mathbf{k}} \,  a^{\dagger}_{\mathbf{k}};
\eea
with the condition $|U_{\mathbf{k}} |^2 \pm |V_{\mathbf{k}} |^2=1$, with $-$ for bosons  and $+$ for fermions, ensuring the canonicity of the transformation.
The transformation (\ref{Atilde}) is conveniently rewritten as
$
\tilde{a}_{\mathbf{k}}(\xi) = J^{-1}(\xi)\,  a_{\mathbf{k}}\,  J(\xi)\, ,
$
where $J(\xi) $  is the generator
which has the property $J^{-1}(\xi)=J(-\xi)$. The transformed operators $\tilde{a}_{\mathbf{k}}(\xi)$ define a state $|\tilde{0}(\xi)\rangle$ through $\tilde{a}_{\mathbf{k}}(\xi)|\tilde{0}(\xi)\rangle=0$, which is related to the vacuum $|0\rangle$ by
$
|\tilde{0}(\xi)\rangle = J^{-1} (\xi)|0\rangle\,.
$
Such a state is a new vacuum of the system, for the following reason,
the above transformation is a unitary operation if $\mathbf{k}$ assumes a \textit{discrete} range of values, i.e. if there is a finite or denumerably infinite number of CCRs. Then the Fock spaces built on the two vacua $|0\rangle$ and $|\tilde{0}(\xi)\rangle$ are equivalent.
%
%
If on the other hand we assume that $\mathbf{k}$ has a continuous infinity of values, which is the situation we are really interested in, we find that the transformation $|\tilde{0}(\xi)\rangle = J^{-1} (\xi)|0\rangle$
%
%
 is not   unitary   any more. This means that the two vacua and thus the two Fock spaces built over them are unitarily inequivalent. We thus have a family of states $|\tilde{0}(\xi) \rangle$, each of which represents in principle a physical vacuum state for the theory. Of course, for these states to be true vacua of the system the issue of stability should be addressed, but this depends on the specific system and is beyond the scope of this simple, free model.

Now, as announced, we shall perform a Bogoliubov transformation on the free Wess--Zumino model and study its effects. The Lagrangian is given by (we adopt the notational conventions of \cite{Wess:1973kz}):
\bea \label{WS}\non
\mathcal{L} &=& \frac{i}{2} \bar{\psi}\gamma_{\mu}\partial^{\mu}\psi + \frac{1}{2}\partial_{\mu}S\partial^{\mu}S + \frac{1}{2}\partial_{\mu}P\partial^{\mu}P
\\ &-& \frac{m}{2}  \bar{\psi}\psi - \frac{m^2}{2} (S^2 + P^2),
\eea
where $\psi$ is a Majorana spinor field, $S$ is a scalar field and $P$ is a pseudoscalar field. This Lagrangian is invariant under the SUSY transformations
\bea
\delta S &=& i{\bar \kappa}\psi\;, \;\; \delta P = i{\bar \kappa} \gamma_5 \psi,\\
\delta \psi &=& \partial_{\mu}(S - \gamma_5 P) \gamma^{\mu}\kappa - m(S + \gamma_5 P)\kappa,
\eea
where $\kappa$ is a Grassmann valued spinorial parameter.

We denote with $\alpha^r_{\mathbf{k}}$, $b_{\mathbf{k}}$ and $c_{\mathbf{k}}$ the annihilators for the fields $\psi$, $S$ and $P$, respectively which annihilate the vacuum $|0\rangle=|0\rangle^{\psi}\otimes|0\rangle^S\otimes|0\rangle^P$
and  we perform simultaneous Bogoliubov transformations on the fermion and on the bosons:
\bea\label{Bog1}
\tilde{\alpha}^r_{\mathbf{k}}(\xi, t) &=& U^{\psi}_{\mathbf{k}}(\xi, t) \, \alpha^r_{\mathbf{k}}(t) + V^{\psi}_{-\mathbf{k}}(\xi, t)  \, \alpha^{r\dagger}_{-\mathbf{k}}( t)\,,
\\[2mm]\label{Bog2}
\tilde{b}_{\mathbf{k}}(\eta,  t) &=& U^{S}_{\mathbf{k}}(\eta,  t) \, b_{\mathbf{k}}(t) - V^{S}_{-\mathbf{k}} (\eta,  t) \,b^{\dagger}_{-\mathbf{k}}(t)\,,
\\[2mm] \label{Bog3}
\tilde{c}_{\mathbf{k}}(\eta,  t) &=& U^{P}_{\mathbf{k}}(\eta,  t) \, c_{\mathbf{k}}(t) - V^{P}_{-\mathbf{k}}(\eta,  t)  \,c^{\dagger}_{-\mathbf{k}}(t)\,.
\eea
The Bogoliubov coefficients of scalar and pseudoscalar bosons are equal each other,  $U^{S}_{\mathbf{k}} =U^{P}_{\mathbf{k}} $ and $V^{S}_{\mathbf{k}} =V^{P}_{\mathbf{k}} $. We thus denote such quantities as $U^{B}_{\mathbf{k}} $ and $V^{B}_{\mathbf{k}} $, respectively.
For fermions an for bosons the Bogoliubov coefficients have the general form:
$
U^{\psi}_{\mathbf{k}} =e^{i\phi_{1\mathbf{k}}}\cos\xi_{\mathbf{k}}(\zeta)$, $ V^{\psi}_{\mathbf{k}} =e^{i\phi_{2\mathbf{k}}}\sin\xi_{\mathbf{k}}(\zeta)$, $
U^{B}_{\mathbf{k}} = e^{i\gamma_{1\mathbf{k}}}\cosh\eta_{\mathbf{k}}(\zeta)$, $ V^{B}_{\mathbf{k}} =e^{i\gamma_{2\mathbf{k}}}\sinh\eta_{\mathbf{k}}(\zeta)
$, respectively, where $\zeta$ represents the relevant parameter which controls the physics underlying the Bogoliubov transformation. For example, $\zeta$ is related to the temperature $T$ in  Thermo Field Dynamics and to the acceleration of the observer in Unruh effect case. Since the phases $\phi_{i\mathbf{k}}$, $\gamma_{i\mathbf{k}}$, with $i=1,2$,  are irrelevant, we neglect them.

The  transformations (\ref{Bog1})--(\ref{Bog3}) can be written at any time $t$ in terms of a generator $J(\xi, \eta, t)$; for examples for fermions we have:
\bea
\tilde{\alpha}^r_{\mathbf{k}}( \xi,  t) &=& J^{-1} (\xi,\eta,  t)\,\alpha^r_{\mathbf{k}}(t) J(\xi,\eta,  t)\,,
\eea
with similar relations holding for the bosonic annihilation and creation operators; in all of them the generator is
$
J(\xi,\eta,   t) = J_{\psi}(\xi,  t) J_{S}(\eta,  t) J_{P}(\eta,  t)\,,
$
where  $J_{\psi}$, $ J_{S}$  and $ J_{P}$ are the generator of the Bogoliubov transformations
for fermion, scalar and pseudoscalar fields \cite{Capolupo:2012vf}.

The new vacuum is
$|\tilde{0}(   t)\rangle=|\tilde{0}(   t)\rangle_{\psi}\otimes |\tilde{0}(  t)\rangle_{S}\otimes |\tilde{0}(   t)\rangle_{P}$, where the states $|\tilde{0}(  t)\rangle_{\alpha}$, with $\alpha = \psi , S , P$, are related to the original ones $|0\rangle_{\alpha}$ by the relations
$
|\tilde{0}(  t)\rangle_{\psi} = J^{-1}_{\psi}(\xi,  t)|0\rangle_{\psi} $,
$ |\tilde{0}(  t)\rangle_{S} = J^{-1}_{S}(\eta,  t)|0\rangle_{S}$,
$  |\tilde{0}(   t)\rangle_{P} = J^{-1}_{P}(\eta,  t)|0\rangle_{P}
$, respectively, therefore the full vacua are related by
\bea
|\tilde{0}(   t)\rangle = J^{-1}(\xi,\eta,  t)|0\rangle\,.
\eea
We notice that $|\tilde{0}(t)\rangle $ has the structure of a condensate of particles, indeed we have
\bea\label{cond1}
 &&\langle\tilde{0}( t)| \alpha^{r\dagger}_{\mathbf{k}} \alpha^r_{\mathbf{k}} |\tilde{0}(  t)\rangle  = |V_{\mathbf{k}}^{\psi}(\xi, t)|^2\,;
\\
&&\langle\tilde{0}(   t)| b^{\dagger}_{\mathbf{k}} b_{\mathbf{k}} |\tilde{0}(   t)\rangle  =  \langle\tilde{0}( t)| c^{\dagger}_{\mathbf{k}} c_{\mathbf{k}} |\tilde{0}( t)\rangle = |V_{\mathbf{k}}^{B}(\eta, t)|^2;\,\,\,\,\,\,\, \label{cond2}
\eea
Such a  structure leads to an energy density different from zero for $|\tilde{0}(t)\rangle $.
To see this explicitly, we must compute the expectation value of the Hamiltonian $H$ corresponding to the Lagrangian in Eq.(\ref{WS}), which has the form $H= H_{\psi} +  H_B $ (where $H_B = H_S + H_P$), on $|\tilde{0}(  t)\rangle$. The results for the two pieces of $H$ are given by
\bea\label{Hpsiasp}
\langle\tilde{0}(  t)| H_{\psi} |\tilde{0}(  t)\rangle =
  - \int \; d^3\mathbf{k}\; \omega_{\mathbf{k}} \,(1- 2 |V^{\psi}_{\mathbf{k}}(\xi, t)|^2)\,,
\eea
and
\bea\label{HBasp}
\langle\tilde{0}(   t)| H_B |\tilde{0}(  t)\rangle = \int\; d^3\mathbf{k}\; \omega_{\mathbf{k}} (1 + 2 |V^{B}_{\textbf{k}}(\eta, t)|^2)\,,
\eea
respectively. We thus obtain the final result
\bea\label{Ht}
\langle\tilde{0}(  t)| H |\tilde{0}(  t)\rangle = 2 \int  d^3\mathbf{k}\; \omega_{\mathbf{k}} (|V^{\psi}_{\textbf{k}}(\xi, t)|^2 + |V^{B}_{\textbf{k}}(\eta, t)|^2)\,
\eea
which is different from zero and positive unless we are in the trivial case $|V^{\psi}_{\textbf{k}}|^2 = |V^{B}_{\textbf{k}}|^2=0$.

The above computation clearly shows that the non zero vacuum condensate energy, and thus the breaking of SUSY, is due to the fact that both the fermion and boson contributions to the condensate lift the vacuum energy by a positive amount, in contrast with the zero point energies which cancel each other.

\section{SUSY breaking and the Anandan-Aharonov invariant}

It has been shown that the presence of the Anandan-Aharonov invariant (AAI) \cite{Anandan:1990fq}, describing the time-energy uncertainty, characterizes the time  evolution of the systems in which the vacuum condensate is physically relevant \cite{Capolupo:2013fasi}.
Then AAIs could be used as a tool to reveal the SUSY breakdown \cite{Capolupo:2013ica}. The AAI appears  in the evolution of any quantum state $|\chi_{\mathbf{k}}(t)\rangle$ which is not stationary, i.e. its  energy uncertainty $\Delta E_{\mathbf{k}} ^{2}(t) = \langle \chi_{\mathbf{k}}(t)|H^{2}|\chi_{\mathbf{k}}(t)\rangle -  (\langle \chi_{\mathbf{k}}(t)|H|\chi_{\mathbf{k}}(t)\rangle)^{2}$  must be non zero. This is the case in the above listed instances \cite{Unruh:1976db}--\cite{Blasone:2002jv}.
When this condition is met, the AAI is defined as (we temporarily restore $\hbar$)
 $S_{\mathbf{k}} = \frac{2}{\hbar} \int_{0}^{  t}  \Delta E_{\mathbf{k}} (t^{\prime}) \, dt^{\prime}\,.$

This invariant is analogous to the geometric phase (but it is defined for non cyclic and non adiabatic evolution) and represents a time-energy uncertainty principle. It can be measured by studying interference of particles or by the analysis of the uncertainty on the
outcome of measurements.

We consider the single particle states
\begin{eqnarray}
|\widetilde{\psi}_{\mathbf{k}}(\xi,  t)\rangle &=&\widetilde{\alpha}^{r\dag}_{\mathbf{k}}(\xi,  t)\rangle|\widetilde{0}(\xi,  t)\rangle_{\psi} = J^{-1}_{\psi}(\xi,  t)|\psi_{\mathbf{k}} \rangle  ,\\
 |\widetilde{S}_{\mathbf{k}}(\eta,  t)\rangle  &=& \widetilde{b}^{\dag}_{\mathbf{k}}(\xi,  t)\rangle|\widetilde{0}(\xi,  t)\rangle_{S}= J^{-1}_{S}(\eta,  t)|S_{\mathbf{k}}\rangle, \\
  |\widetilde{P}_{\mathbf{k}}(\eta,   t)\rangle &=& \widetilde{c}^{\dag}_{\mathbf{k}}(\eta,  t)\rangle|\widetilde{0}(\eta,  t)\rangle_{P}= J^{-1}_{P}(\eta,  t)|P_{\mathbf{k}}\rangle.
\end{eqnarray}
  The energy variances of these states  are $\Delta E_{\mathbf{k}}^B(t) = \sqrt{2} \omega_{\bf k}|U^B_{\bf k}(\eta,t)| |V^B_{\bf k}(\eta,t)|$  and
$\Delta E_{\mathbf{k}}^{\psi}(t) =   \omega_{\bf k} |U^\psi_{\bf k}(\eta,t)| |V^\psi_{\bf k}(\eta,t)|,$ respectively. Then the corresponding AAIs are given by
\bea\label{AAI-Bog-Bos}
 S_{\mathbf{k}}^S(t) =S_{\mathbf{k}}^{P}(t)=2\,\sqrt{2}\int_{0}^{  t} \omega_{\bf k} |U^B_{\bf k}(\eta,t^{\prime})|| V^B_{\bf k} (\eta,t^{\prime})|\,dt^{\prime}
\eea
for scalar and pseudoscalar bosons  and
\bea\label{AAI-Bog-Fer}
S_{\mathbf{k}}^{\psi}(t)\,=\,2 \int_{0}^{  t}\omega_{\bf k}|U^\psi_{\bf k}(\xi,t^{\prime})| |V^\psi_{\bf k}(\xi,t^{\prime})|\, dt^{\prime},
\eea
for the Majorana fermion field. Such invariants signal the presence of the condensate, since their
 values are controlled by the Bogoliubov coefficients and they vanish as the condensates disappear, i.e.
 when $U_{\bf k}$ and $V_{\bf k}$ are zero\footnote{We notice that the particle mixing phenomenon is peculiar for the following reason. Although also in this case SUSY is spontaneously broken by a condensate \cite{Capolupo:2010ek,Mavromatos:2010ni}, in this case the AAI arises mainly as an effect of the mixing of fields with only a small contribution due to the condensate structure \cite{Blasone:2009xk,Capolupo:2011rd}. Therefore, in this case the presence of the AAI is not directly linked with the presence of the condensate. In all the other cases instead the AAI is entirely due to the condensate contribution.}.

Now we study the specific case of thermal states and propose a possible experiment to detect thermal SUSY violation by measuring nonvanishing AAIs. As is clear from all we said, in the TFD formalism \cite{Takahashi:1974zn} the thermal vacuum is a condensate generated through Bogoliubov transformations whose parameter is related to temperature. The Bogoliubov coefficients $U$ and $V$ have the general form \cite{Takahashi:1974zn}
$U_{{\bf k}} = \sqrt{\frac{e^{\beta \omega_{\bf k} }}{e^{\beta \omega_{\bf k } }\pm 1}}$ and
$V_{{\bf k}} = \sqrt{\frac{1}{e^{\beta \omega_{\bf k} } \pm 1}}$, with $-$ for bosons and $+$ for fermions, and $\beta = 1/ k_{B}T$.

 The energy variances of a temperature dependent single particle state are given by
\bea\non
\Delta E_{\mathbf{k}}^{S}  &=& \Delta E_{\mathbf{k}}^{P}\nonumber= \sqrt{2}\, \omega_{\bf k} \, U^{B}_{\bf k} V^{B}_{\bf k} = \\ &=&  \sqrt{2} \omega_{\bf k} \,\frac{{e^{\beta \omega_{\bf k}/2  }}}{{(e^{\beta \hbar \omega_{\bf k } }-1)}}\,,
\eea
for the bosonic states, and
\bea\non
\Delta E_{\mathbf{k}}^{\psi} =  \omega_{\bf k}^{\psi} \, U^{\psi}_{\bf k} V^{\psi}_{\bf k}
=  \hbar \omega_{\bf k}^{\psi} \,\frac{{e^{\beta  \omega_{\bf k}^{\psi}/2  }}}{{(e^{\beta  \omega_{\bf k }^{\psi} }+1)}}\,,
\eea
for the fermionic state.
 The corresponding AAIs are
\bea\label{ThermalAAIs}
S_{\mathbf{k}}^{S}&=& S_{\mathbf{k}}^{P}= 2\sqrt{2}\, \omega_{\bf k}\, t\, \frac{e^{ \beta\hbar \omega_{\bf k}/2  }}{e^{\beta\hbar \omega_{\bf k } } - 1}\,,\nonumber
\\
S_{\mathbf{k}}^{\psi} &=& 2 \,\omega_{\bf k}\, t\, \frac{e^{ \beta\hbar \omega_{\bf k}/2  }}{e^{\beta\hbar \omega_{\bf k } } + 1}\,.
\eea
In a supersymmetric model, at $T\neq 0$, the above invariants are different from zero.

\section{Experimental realization}

The presence of the AAIs and then the SUSY violation could be tested by employing a mixture of cold fermion  atoms and diatomic molecules trapped in two dimensional optical lattices \cite{Yue Yang}, in which the Wess-Zumino model in $2+1$ dimensions can emerge at low energies. Such a system displays Dirac points in the Brillouin zone, therefore the excitations will have relativistic dispersion relations and SUSY will be described by the super-Poincar\'{e} algebra, in contrast with other setups proposed in the literature, which display a nonrelativistic version of SUSY. The superpartner of the fermionic atom is a bosonic diatomic molecule. The setup allows to simulate both the massless and the massive models, the latter being attained by putting a Bose-Einstein condensate of dimolecules nearby, allowing exchange of pairs of molecules with the mixture through Josephson tunneling \cite{Yue Yang}.

SUSY breaking is expected when the system is put at nonzero temperature. A proof of this breaking can come from the detection a thermal Goldstone fermion, the phonino, which is predicted to appear in this case \cite{Midor,Midor2,Midor3}. This would present experimental difficulties, however an alternative signal can come from the detection of the vacuum condensate. Building on \cite{Yue Yang},
 we propose to measure  the difference between the geometric invariants in the evolution of two mixtures of cold atoms and molecules trapped in two coplanar, two-dimensional optical lattices, one at temperature $T\neq 0$ and the other one at $T = 0$ \cite{Capolupo:2013ica}.
 Excitations of the lattice at $T\neq 0$ will be associated to the non-vanishing AAI invariants (\ref{ThermalAAIs}), while those in the $T=0$ lattice, on the other hand, have vanishing AAI. The presence of the condensate introduce a nontrivial modification on the uncertainty, therefore resulting an inevitably increase of the uncertainty on the
outcome of measurements. Then, a study of the AAIs can be done also by analyzing the energy uncertainty in the lattice at $T\neq 0$.

Non trivial values of the AAIs as functions of temperature are obtained (see In Figs. 1 and 2), by considering temperatures of the order of $(20-200) nK$, atomic excitation frequencies characteristic  of Bose-Einstein condensates, i.e. $\omega$ of order of $ 2 \times 10^{4} s^{-1}-10^{5} s^{-1} $, and time intervals of order of $t =  1 /\omega$ \cite{Capolupo:2013ica}. The values of the AAI we found are in principle detectable.
\begin{figure}
\begin{picture}(300,180)(0,0)
\put(10,20){\resizebox{8.5 cm}{!}{\includegraphics{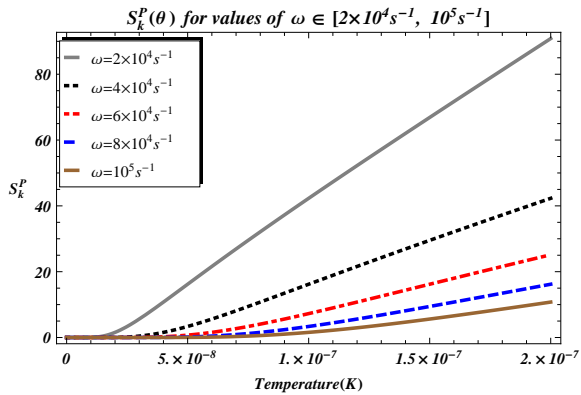}}}
\end{picture}\vspace{-1cm}
\caption{\em Plots of AAI for bosons, $S_{\mathbf{k}}^{S,P}$
as a function of temperature $T$, for a time interval $t = 1 /\omega $ and for sample values of  $\omega \in  [2 \times 10^{4} s^{-1},  10^{5} s^{-1}]$;  $\omega =  2 \times 10^{4} s^{-1}$  (gray solid line),   $\omega =
4 \times 10^{4} s^{-1}$ (black dotted line), $\omega = 6 \times 10^{4} s^{-1}$ (red dot-dashed line),
$\omega = 8 \times 10^{4} s^{-1}$ (blue dashed line), $\omega = 10^{5} s^{-1}$  (brown solid line).}
\label{pdf}
\end{figure}
\begin{figure}
\begin{picture}(300,180)(0,0)
\put(10,20){\resizebox{8.5 cm}{!}{\includegraphics{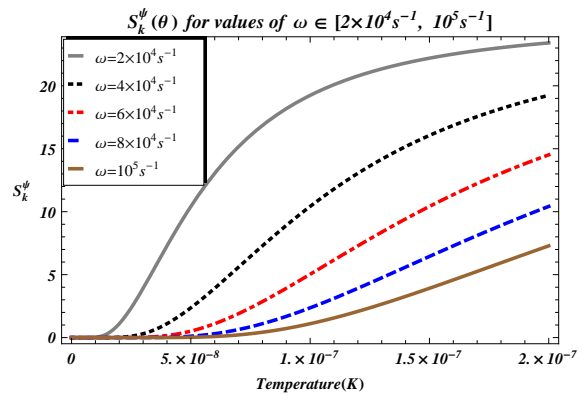}}}
\end{picture}\vspace{-1cm}
\caption{\em Plots of AAI for fermions, $S_{\mathbf{k}}^{\psi}$
as a function of temperature $T$, for the same time interval and sample values of  $\omega$   as  in Figure 1.}
\label{pdf}
\end{figure}

At temperatures above $\thickapprox 200 ~nK$, the condensate (and thus the AAIs) is expected to disappear.
As a final comment, we notice that, as happens for any system which presents a condensate structure~\cite{Capolupo:2013fasi}, also in the present context AAIs are unaffected by the presence of noise.

In conclusion, we have shown that, in the free Wess-Zumino model, all the phenomena characterized by the presence of the vacuum condensate generate a spontaneous SUSY breaking  due to the non zero vacuum energy. Indeed,  bosons and fermion condensates both lift the vacuum energy by a positive amount.  Such a breaking could be detected by measuring the AAIs generated by the condensates in a thermal bath  in an optical lattice simulating the Wess-Zumino model.

\section*{Acknowledgements}
Partial financial support from MIUR and INFN is acknowledged.

\section*{Conflict of Interests}
The authors declares that there is no conflict of interests regarding the publication of this article.


\begin{thebibliography}{999}

\bibitem{Golfand:1971iw}
  Y.~A.~Golfand and E.~P.~Likhtman,
  JETP Lett.\  {\bf 13} (1971) 323
   [Pisma Zh.\ Eksp.\ Teor.\ Fiz.\  {\bf 13} (1971) 452].

\bibitem{Wess:1992cp}
  J.~Wess and J.~Bagger,
  \emph{Supersymmetry and supergravity},
  Princeton, USA: Univ. Pr. (1992) 259 p

\bibitem{Weinberg:1976zw}
  S.~Weinberg,
  Phys.\ Lett.\  B {\bf 62} (1976) 111.



\bibitem{Witten:1981nf}
  E.~Witten,
 {\it   Nucl.\ Phys.\  B} {\bf 188}, 513 (1981).

\bibitem{Shadmi:1999jy}
  Y.~Shadmi, Y.~Shirman,
 {\it   Rev.\ Mod.\ Phys.\ } {\bf 72}, 25 (2000).



\bibitem{Yue Yang}
Y.~Yu, K.~Yang
{\it    Phys. Rev. Lett.} {\bf 105}, 150605 (2010).

\bibitem{Capolupo:2012vf}
  A.~Capolupo and M.~Di Mauro,
  {\it Phys.\ Lett.\ A}, {\bf   376},  2830 (2012);

 \bibitem{Capolupo:2013af}
  A.~Capolupo and M.~Di Mauro,
  {\it   Acta Phys.\ Polon.\ B} {\bf 44}, 81 (2013).

\bibitem{Takahashi:1974zn}
  Y.~Takahashi, H.~Umezawa,
{\it    Collect.\ Phenom.\ } {\bf 2}, 55 (1975); reprinted in Int. J. Mod. Phys. B {\bf 10}, 1755 (1996);





\bibitem{Schwinger:1951nm}
  J.~S.~Schwinger,
 {\it   Phys.\ Rev.\ } {\bf 82}, 664 (1951).

\bibitem{Unruh:1976db}
  W.~G.~Unruh,
{\it    Phys.\ Rev.\  D} {\bf 14}, 870 (1976).

\bibitem{Bardeen:1957mv}
  J.~Bardeen, L.~N.~Cooper, J.~R.~Schrieffer,
 {\it   Phys.\ Rev.\   } {\bf 108}, 1175 (1957).

\bibitem{Iorio:2010pv}
  A.~Iorio,
 {\it   Annals Phys.\ } {\bf 326}, 1334 (2011).
\bibitem{Blasone:2001du}
  M.~Blasone, A.~Capolupo, O.~Romei, G.~Vitiello,
{\it    Phys.\ Rev.\ D} {\bf 63}, 125015 (2001).

\bibitem{Blasone:2002jv}
  M.~Blasone, A.~Capolupo, G.~Vitiello,
 {\it   Phys.\ Rev.\ D } {\bf 66}, 025033 (2002).

\bibitem{Buchholz:1997mf}
  D.~Buchholz, I.~Ojima,
{\it    Nucl.\ Phys.\  B} {\bf 498}, 228 (1997).

\bibitem{Das:1997gg}
  A.~K.~Das, Finite temperature field theory,
World Scientific (Singapore) 1997.




\bibitem{Bogolyubov:1947zz}
  N.~N.~Bogolyubov,
  J.\ Phys.\ (USSR) {\bf 11} (1947) 23
   [Izv.\ Akad.\ Nauk Ser.\ Fiz.\  {\bf 11} (1947) 77].

\bibitem{Umezawa:1993yq}
  H.~Umezawa,
  \emph{Advanced field theory: Micro, macro, and thermal physics},
New York, USA: AIP 1993.


%
%
\bibitem{Wess:1973kz}
  J.~Wess, B.~Zumino,
 {\it   Phys.\ Lett.\  B} {\bf 49}, 52 (1974).
\bibitem{Capolupo:2010ek}
  A.~Capolupo, M.~Di Mauro, A.~Iorio,
 {\it   Phys.\ Lett.\  A} {\bf 375}, 3415 (2011).


\bibitem{Mavromatos:2010ni}
  N.~E.~Mavromatos, S.~Sarkar, W.~Tarantino,
 {\it   Phys.\ Rev.\  D} {\bf 84}, 044050 (2011).


\bibitem{Capolupo:2013ica}
  A.~Capolupo and G.~Vitiello,
  Adv.\ High Energy Phys.\  {\bf 2013} (2013) 850395

\bibitem{Anandan:1990fq}
  J.~Anandan and Y.~Aharonov,
{\it   Phys.\ Rev.\ Lett.}  {\bf 65}, 1697 (1990).

\bibitem{Capolupo:2013fasi}
  A.~Capolupo, G.~Vitiello,
 {\it   Phys.\ Rev.\  D} {\bf  88}, 024027 (2013).


\bibitem{Capolupo:2011rd}
  A.~Capolupo,
 {\it     Phys.\ Rev.\ D} {\bf 84}, 116002 (2011).



\bibitem{Blasone:2009xk}
M.~Blasone, A.~Capolupo, E.~Celeghini, G.~Vitiello,
 {\it  Phys.\ Lett.\ B} {\bf 674}, 73 (2009).

\bibitem{Midor}
S.~Midorikawa, {\it  Prog. Theor. Phys.} {\bf 73}, 1245 (1985).

\bibitem{Midor2}
V.~V.~Lebedev and A.~V.~Smilga, {\it  Nucl. Phys. B} {\bf 318}, 669 (1989).

\bibitem{Midor3}
K.~Kratzer, {\it  Ann. Phys.} (N.Y) {\bf 308}, 285 (2003).


\end{thebibliography}
\end{document}